\documentclass[a4paper,11pt]{article}
\usepackage{jcappub} 
\usepackage{lineno}
\usepackage{multirow}
\usepackage{nasa}

\arxivnumber{2401.12403} 
\title{Multi-energy diffuse neutrino fluxes originating from core-collapse supernovae}

\author{Yosuke Ashida}
\affiliation{Department of Physics, Tohoku University,\\
Sendai, Miyagi 980-8578, Japan}

\emailAdd{yosuke.ashida.a1@tohoku.ac.jp}

\abstract{
Diffuse neutrino fluxes attributed to two different physical processes in core collapse of massive stars are visited with their potentiality of exploring stellar physics more deeply being stressed.
In this work, available models of thermal MeV-scale neutrinos produced at the core of collapsing stars and non-thermal high-energy neutrinos emitted from accelerated cosmic rays interacting with circumstellar material are bridged through features of core-collapse supernovae such as progenitor mass and optical properties.
The calculated diffuse fluxes are presented with discussion about their detection prospects at neutrino telescopes. 
}

\begin{document}
\maketitle
\flushbottom

\section{Introduction} \label{sec:intro}

Massive stars are known to undergo a fierce fate at the end of their life; their self-gravity surpasses pressures with nuclear fusion at the final stage of stellar evolution and the following shock wave produced by the nuclear repulsion force (core bounce) propagates outwards to blow off the stellar envelope (e.g., see Refs.~\cite{2006RPPh...69..971K,2012ARNPS..62..407J} for more details). 
This is called core-collapse supernovae (CCSNe) and expected for stars with masses of $\gtrsim 8 M_\odot$.
Throughout the core-collapse process, {\it thermal} neutrinos of all flavors are expected to be produced, e.g., via electron capture and pair annihilation (see Refs.~\cite{2013RvMP...85..245B,2018SSRv..214...31T,2020LRCA....6....3M,2021Natur.589...29B} for recent reviews). 
Such neutrinos are believed to be a key for successful explosions as they could prevent momentum of the shock wave from diluting before reaching the stellar surface.
These thermal neutrinos from the star's core are emitted up to a timescale of $O(10)$~s from the core bounce and their typical energy scale is $O$(1--10)~MeV~\cite{2012ApJ...749...98T,2013ApJS..205....2N,2019ApJ...881..139S,2021PTEP.2021a3E01S}. 
Although there have been huge efforts made on the theoretical side for deeper understanding of the CCSN mechanism both numerically and analytically, a complete picture of the collapsing process is still awaited. 
On the experimental side, there has been only one successful observation of supernova MeV neutrinos, which is SN~1987A in the Large Magellanic Cloud~\cite{1987JETPL..45..589A,1987PhRvL..58.1490H,1987PhRvL..58.1494B}. 

In recent years, further neutrino emission from CCSNe at later times with a much longer timescale, $O(1)$~hour to $O(1)$~year, due to confined dense circumstellar material (CSM) is actively studied~\cite{2011PhRvD..84d3003M,2016APh....78...28Z,2017MNRAS.470.1881P,2018PhRvD..97h1301M,2019ApJ...872..157W,2019SCPMA..6259511L,2022JCAP...08..011S,2023PhRvD.108j3033S,2024PhRvD.109j3020M,2025ApJ...982...93S}. 
CSM is predicted to originate from mass loss that massive stars experience 0.1--100~years ahead of core collapse~\cite{2014ARA&A..52..487S,2014MNRAS.439.2917M,2018MNRAS.476.2840M,2021ApJ...914...64T} and its existence is strongly supported by recent optical observations of supernovae~\cite{2017hsn..book..403S,2017NatPh..13..510Y,2018ApJ...858...15M,2018ApJ...861...63H,2018NatAs...2..808F,2020MNRAS.496.1325B,2021ApJ...923L..24S,2022MNRAS.512.2777T,2022hxga.book...75M}. 
Likely processes that cause the star's mass loss are, for example, stellar winds and interactions between binary stars~\cite{2012ARA&A..50..107L,2015ApJ...805...20S}. 
The supernova ejecta interacts with the surrounding CSM, accelerating charged particles therein to relativistic energies via the Fermi shock acceleration process~\cite{1997ApJ...490..619S,2014BrJPh..44..415B,2016crpp.book.....G}. 
These relativistic particles then interact with non-relativistic particles ({\it pp} collisions) to produce charged mesons that decay into neutrinos, e.g., $pp \rightarrow \pi^+ \rightarrow \mu^+ \nu_\mu \rightarrow e^+ \nu_e \bar{\nu}_\mu \nu_\mu$. 
Energies of neutrinos from such {\it non-thermal} process are expected to reach $O$($10^4$--$10^6$)~GeV depending on the CSM feature and so on. 
Emission timescale, $O(1)$~hour to $O(1)$~year, is determined by the ejecta velocity and CSM density profile.
There are three reported SNe, time and location of whose observations in optical survey coincide with IceCube-detected high-energy neutrino events~\cite{2025arXiv250808355S,2026ApJ...997..145L}.
A strong statement is too early at the moment because of the limited statistics; however, the enhanced optical survey and neutrino observation expect to provide more opportunities on such sources~\cite{2023ApJ...945...98V,2023ApJ...956L...8K,2024PhRvL.132f1001W}.

Measuring the accumulated flux of supernova neutrinos released over the cosmic history serves another direction of probing the supernova feature. 
In contrast to detecting neutrinos from a transient source, such methodology is useful for acquiring more general and average picture and holds a merit of not requiring to wait for another nearby supernova. 
The integrated flux of thermal neutrinos is referred to as the diffuse supernova neutrino background (DSNB), or supernova relic neutrinos (SRNs) in some articles, and still remains undiscovered in spite of intensive searches at neutrino detectors around the world, such as Super-Kamiokande and KamLAND~\cite{2021PhRvD.104l2002A,2023ApJ...951L..27H,2025arXiv251102222A,2022ApJ...925...14A} (e.g., Refs.~\cite{2022arXiv220709632S,2023PJAB...99..460A} provide reviews of recent progress in theoretical and experimental DSNB studies). 
In the high-energy regime, the diffuse neutrino flux of astrophysical-origin beyond the known atmospheric background has been measured by IceCube~\cite{2015PhRvD..91b2001A,2019PhRvD..99c2004A,2021PhRvD.104b2002A,2022ApJ...928...50A}. 
Their origin is not completely ascertained; however, observations of high-energy neutrinos in the direction of some astronomical objects are reported, from the blazar TXS~0506+056~\cite{2018Sci...361..147I} and the Seyfert galaxy~\cite{2022Sci...378..538I,2025arXiv251013403A}. 
Neutrinos from CCSNe through the inelastic {\it pp} process mentioned above may contribute partially to the observed diffuse flux~\cite{2016APh....78...28Z,2017MNRAS.470.1881P,2018PhRvD..97h1301M,2019ApJ...872..157W,2019SCPMA..6259511L,2022JCAP...08..011S}. 

So far, theoretical studies on neutrino emissions from the core and outer region of CCSNe have been made mostly independently, as tracing physics over the entire dynamics throughout core collapse to ejecta-CSM interactions is quite challenging.
Nevertheless, utilizing neutrinos from those two processes by any means should confer benefits for multifaceted comprehension of CCSNe. 
For example, Ref.~\cite{2025ApJ...982...93S} proposes a potential opportunity of investigating the origin of CSM through simultaneous detection of low-energy pre-SN neutrinos and high-energy neutrinos from a CCSN at different detectors.
Considering the global situation that many different neutrino telescopes start operating or are being built in the upcoming years, studies on such multi-energy neutrino astronomy become more valuable.
In particular, this paper focuses on the diffuse neutrino measurement as such studies are still missing in that area.
In this work, because there is no complete framework that consistently treats both core and outer regions, available models on each region are referred and forcibly linked based on the observationally-motivated assumptions. 
For this purpose, a conventional classification of supernovae based on observed lines and spectral features in the optical survey~\cite{2003LNP...598...21T,2017hsn..book..195G} is referred in conjunction with progenitor features, particularly their initial mass. 

The rest of this paper is structured as follows. 
Section~\ref{sec:snclass} introduces the supernova classification used for the later calculation of the diffuse fluxes and is followed by Section~\ref{sec:nuemiss} in which models of thermal MeV neutrinos and non-thermal high-energy neutrinos are described. 
The calculated neutrino fluxes are presented in Section~\ref{sec:nuflux} and the results are discussed in more detail in Section~\ref{sec:discuss}. 
Finally, concluding remarks are given in Section~\ref{sec:concl}.

\section{Supernova classification} \label{sec:snclass}

Supernovae (SNe) are caused either through a thermonuclear or core-collapse process depending on their mass. 
Apart from that, they are classified based on their light curves and spectral features in the optical survey~\cite{2003LNP...598...21T,2011MNRAS.412.1441L,2017hsn..book..195G}. 
The presence of silicon lines indicates a thermonuclear process and SNe with such lines are labeled as Type Ia. 
The others are considered to be based on a core-collapse process. 
Among them, SNe without hydrogen lines are Type I which are then classified as Type Ib (Ic) if the helium lines are (not) observed. 
SNe with hydrogen lines are assorted into Type II and further subdivided based on the observed optical properties into II-L, II-P, IIn and IIb. 
SNe Type II-L show a light curve with a steady (``L''inear) decline along the time, while SNe Type II-P retain their brightness at later times (``P''lateau). 
SNe Type IIn show narrow or intermediate hydrogen emission lines (``N''arrow). 
SNe Type IIb have weak hydrogen lines and then resemble Type Ib which do not show hydrogen lines. 
In the following part of the paper, neutrino emission from CCSNe, i.e., SNe other than Type Ia, is focused. 

  \begin{table}[htbp]
  \begin{center}
  \caption{SN classification, portion, original situation, and progenitor model assumed in the present work.}
  \label{tab:snclass}
  \vspace{+1truept}
    \begin{tabular}{c c c c} \hline 
      SN type & fraction & origin & progenitor model \\ \hline 
      II-P & 48.2\% & single & CNS \\ 
      II-L &  6.4\% & single & CNS \\
      IIn  &  8.8\% & single & HNS \\
      IIb  & 10.6\% & binary & CNS, HNS \\
      Ib   & \multirow{2}{*}{26.0\%} & binary & CNS, HNS \\
      Ic   &  & single, binary & CNS, HNS \\ \hline
    \end{tabular}
  \end{center}
  \end{table}

The observed lines and spectral shapes should reflect stellar envelope and circumstellar environment which are expected to be the outcomes of stellar evolution. 
Among important factors that determine evolutionary path are star's mass, metallicity, stellar winds, and binary interactions. 
For example, SNe Type Ibc (Ib and Ic) are expected to be heavier than Type II, if they are single without a companion star, because of active stripping of the envelope~\cite{2011MNRAS.412.1522S}. 
SNe Type Ic are considered to be heavier than Ib for the same reason. 
It should be noted that this may differ for binary stars as mass transfer makes the situation more complex~\cite{2011MNRAS.412.1522S,2021PhRvD.103d3003H,2024MNRAS.532.3926K}.
The narrow hydrogen lines observed from SNe Type IIn imply active interactions between the SN ejecta and CSM. 
In Ref.~\cite{2011MNRAS.412.1522S}, multiple cases are considered to subdivide the initial mass function (IMF) so that the resulting fractions of CCSN types are consistent with the observation in the Lick Observatory Supernova Search (LOSS). 
Here, the Salpeter IMF ($\psi_{\rm IMF} \propto M^{-2.35}$)~\cite{1955ApJ...121..161S} is adopted and the mass range of $8.5$ to $150 M_\odot$ is assumed for CCSNe. 
Among multiple scenarios proposed in Ref.~\cite{2011MNRAS.412.1522S}, the ``hybrid \#1'' scenario (see Figure~7 in that paper) is adopted for the current study. 
In this scenario, Type II-P, II-L, IIn, and roughly half of Ic are considered to arise from single stars and IIb, Ib, and the remainder of Ic result via binary systems. 
This scenario has several strengths while no obvious disadvantages. 
The assigned mass range for II-P ($8.5$--$18.7 M_\odot$) is in a good agreement with the observation and the narrow range for II-L ($18.7$--$23.1 M_\odot$) is also consistent. 
Most massive single stars die either as IIn ($23.1$--$37 M_\odot$) or Ic ($37$--$150 M_\odot$). 
These different fates may be determined by other factors that change the mass-loss effieincy of single stars, such as rotation and metallicity (e.g., Ref.~\cite{2011MNRAS.412.1441L} reports that SNe Type IIn tend to be observed in lower-metallicity galaxies).
The binary star case (IIb, Ib, and partial Ic) is sampled over the entire mass range, $8.5$--$150 M_\odot$, as the mass-loss efficiency is determined not only by the initial mass but also by other factors such as the binary separation for binary stars. 
This gives a good explanation for the tiny observed difference between Ib and IIb. 
For calculating the DSNB flux in Section~\ref{sec:nuflux}, the progenitor model has to be selected for each SN type. 
As described in Section~\ref{subsec:mevproc}, models of a progenitor with $15 M_\odot$ (CNS) and $40 M_\odot$ (HNS) are used in this study, and either model is assigned to each SN type. 
The CNS (HNS) model is used for SNe that are mapped as lighter (heavier) than $23.1 M_\odot$ on the Salpeter IMF, which constitutes 75.7\% (24.3\%) of total. 
The consequent fraction of high-mass neutron stars (24.3\%) is consistent with the fraction from observations of binary pulsar systems ($\sim$20\%)~\cite{2016arXiv160501665A}. 
This choice is supported also by other studies that report SNe Type IIn and Ic are likely to result from very massive stars~\cite{2007ApJ...656..372G,2011ApJ...732...63S,2018MNRAS.480.2072K}, although the potential systematic uncertainty should be noted. 
The SN type, fraction, original situation, and corresponding progenitor model in this study are summarized in Table~\ref{tab:snclass}.

\section{Neutrino emission} \label{sec:nuemiss}

\subsection{Neutrinos from thermal process at stellar core} \label{subsec:mevproc}

There are majorly two fates after the star's core collapse; a {\it neutron star} (NS) is left after a successful blow off of the envelope and cooling of the core (successful SN explosion), or a {\it black hole} (BH) is formed as continuous mass accretion leads to the critical point (failed SN).\footnote{Fallback SNe that undergo successful explosions but lead to the BH formation because of the late-time substantial fallback may exist but are considered to be rare.}
Neutrinos are expected to be emitted from both cases and their energy spectra are known to differ~\cite{2013ApJS..205....2N,2021PASJ...73..639N}. 
However, considering the BH formation requires a modification of the mapping on the Salpeter IMF adopted from Ref.~\cite{2011MNRAS.412.1522S} (see Section~\ref{sec:snclass}). 
In addition, the fraction of such BH formation case as a function of progenitor mass is not well known. 
For these reasons, contributions from failed SNe are ignored in the main part of this paper, which does not change the scope of this study, and relevant discussion is given later in Section~\ref{sec:discuss}. 
As a model of neutrino emission from the successful explosion that forms a NS, the numerical simulation results from a previous work~\cite{2022ApJ...925...98N} are adopted, in which two cases on the remaining NS mass (baryonic NS masses of $1.47 M_\odot$ and $1.86 M_\odot$) are calculated.
The corresponding progenitors for these two cases are taken from Ref.~\cite{1995ApJS..101..181W} and have the solar metallicity with zero-age-main-sequence (ZAMS) masses of $15 M_\odot$ and $40 M_\odot$, respectively. 
In the following part, these two models are referred to as the canonical-mass NS (CNS) and high-mass NS (HNS), respectively.
As summarized in Table~\ref{tab:snclass}, the CNS and HNS models are assigned to SNe mapped as lighter and heavier than $23.1 M_\odot$ on the Salpeter IMF, respectively. 

The neutrino emission profile (total and average energies of emitted neutrinos) depends strongly on the choice of nuclear equation of state (EOS)~\cite{2018PhRvC..97c5804N,2019ApJ...878...25N,2020ApJ...891..156N,2022ApJ...925...98N}. 
In the present work, three types of nuclear EOS are considered: LS220 EOS~\cite{1991NuPhA.535..331L}, Shen EOS~\cite{2011ApJS..197...20S}, and Togashi EOS~\cite{2017NuPhA.961...78T}.  
Properties of the remnant NS and the resulting neutrino signal for each EOS case are summarized in Table~1 of Ref.~\cite{2022ApJ...937...30A}. 
Detailed neutrino light curves for each case are available in Refs.~\cite{2021PASJ...73..639N,2022ApJ...925...98N}. 
In a brief summary, the emitted neutrino amount is largest (smallest) with Togashi (Shen) EOS for both CNS and HNS cases.

\subsection{Neutrinos from {\it pp} interactions in outer region}

In this study, models from Refs.~\cite{2018PhRvD..97h1301M,2024PhRvD.109j3020M} are adopted for neutrino production from ejecta-CSM interactions. 
The models are characterized by the CSM density profile, $D r^{-2}$, where $r$ is distance and $D \equiv 5 \times 10^{16} \ {\rm g \ cm^{-1}} D_*$ ($D$ is related to the mass-loss rate and wind velocity). 
$D_*$ differs for each SN type as summarized in Table~1 of Ref.~\cite{2018PhRvD..97h1301M}. 
The most dense CSM ($D_* = 1$) is assumed for SNe Type IIn, while smaller $D_*$'s are used for the other types ($D_* = 10^{-2}$ for II-P, $10^{-3}$ for II-L/IIb, and $10^{-5}$ for Ibc). 
Here, for II-P, the case with an enhanced CSM motivated by SN~2013fs~\cite{2017NatPh..13..510Y} is used. 

Cosmic rays are accelerated by the shock acceleration mechanism similarly to the SN remnants~\cite{2021Univ....7..324C}. 
Such cosmic rays then interact with low-energy nucleons in CSM to produce mesons which are parents of high-energy neutrinos. 
Here, the cosmic ray spectrum following a power law index of $s$ is considered. 
The resulting neutrino energies are typically $\sim$5\% of the parent proton energy~\cite{2006PhRvD..74c4018K}. 
All neutrino flavors ($\nu_e$, $\bar{\nu}_e$, $\nu_\mu$, $\bar{\nu}_\mu$, $\nu_\tau$, $\bar{\nu}_\tau$) are produced equally in the current scenario. 
The timescale that neutrino emission begins depends on the SN type; $O(1)$~hour for Ibc, II-L, and IIb, $O(1)$~day for II-P, and $O(10)$~day for IIn.
Among different SNe, Type IIn gives the largest neutrino flux as its CSM density is highest, and followed by II-P, II-L/IIb, and Ibc in the descending order.  
Neutrino light curves and fluxes from each SN type are shown in Refs.~\cite{2018PhRvD..97h1301M,2024PhRvD.109j3020M} (e.g., see Figure~2 of Ref.~\cite{2024PhRvD.109j3020M}).

\section{Diffuse neutrino flux} \label{sec:nuflux}

\subsection{Diffuse flux of thermal SN neutrinos (DSNB)} \label{subsec:dsnb}

The DSNB flux is calculated by integrating thermal neutrinos from CCSNe over the cosmic history as: 
  \begin{eqnarray}
    \frac{d\phi_{\rm DSNB}(E_\nu)}{dE_\nu}=c\int^{z_{\rm max}}_{0} R_{\rm CC}(z) \left\langle \frac{dN_{\rm thermal}(E^\prime_\nu)}{dE^\prime_\nu} \right\rangle \times \frac{dz}{H_0\sqrt{\Omega_{\rm m}(1+z)^3+\Omega_\Lambda}},
  \label{eq:dsnbflux}
  \end{eqnarray}
where $c$ is the speed of light and $\Omega_{\rm m}= 0.2726$, $\Omega_\Lambda= 0.7274$, and $H_0= 70.4$~km~s$^{-1}$ are the cosmological constants~\cite{2014MNRAS.444.1518V}. 
The neutrino energy at Earth ($E_\nu$) is related to that at a SN ($E^\prime_\nu$) with a redshift $z$ as $E^\prime_\nu = (1+z) E_\nu$.
The integration is performed up to $z_{\rm max}=5$ in this study as contributions from $z>5$ are tiny.  
The average neutrino number spectrum is obtained by mixing number spectra for each fate ($dN_{\rm CNS}(E^\prime_\nu)/dE^\prime_\nu$ and $dN_{\rm HNS}(E^\prime_\nu)/dE^\prime_\nu$) with a fraction of high-mass neutron star cases to all CCSNe ($f_{\rm HNS}$) as: 
  \begin{eqnarray}
    \left\langle \frac{dN_{\rm thermal}(E^\prime_\nu)}{dE^\prime_\nu} \right\rangle = f_{\rm HNS}\frac{dN_{\rm HNS}(E^\prime_\nu)}{dE^\prime_\nu} + (1-f_{\rm HNS})\frac{dN_{\rm CNS}(E^\prime_\nu)}{dE^\prime_\nu}. 
  \label{eq:nthermonu}
  \end{eqnarray}
The core-collapse rate $R_{\rm CC}(z)$ is calculated by utilizing the Illustris-1 cosmological simulation~\cite{2014MNRAS.444.1518V,2015A&C....13...12N} as: 
  \begin{eqnarray}
    R_{\rm CC}(z)=\dot\rho_\ast(z) \frac{\int^{M_{\rm max}}_{M_{\rm min}} \psi_{\rm IMF}(M)dM}{\int^{M_{\rm max}}_{0.1M_\odot} M \psi_{\rm IMF}(M)dM},
  \label{eq:ccrate}
  \end{eqnarray}
where $\dot\rho_\ast(z)$ is the cosmic star formation rate density (SFRD). 
Here, $M_{\rm min}=8.5M_\odot$ and $M_{\rm max}=150M_\odot$ and the Chabrier IMF~\cite{2003PASP..115..763C}, that follows a lognormal function for $M < 1M_\odot$ and a power law with the same slope as the Salpeter IMF ($-2.35$) for $M > 1M_\odot$, are used. 
Note that the results are not affected by a different IMF choice between the SN classification in Section~\ref{sec:snclass} and Eq.~(\ref{eq:ccrate}) as both the Salpeter and Chabrier IMFs are consistent in the mass range for CCSNe ($M > 8.5M_\odot$ in this study). 
In the present work, the core-collapse rate equals the rate of CCSNe (successful explosions) as the BH formation case is not considered. 

\begin{figure}[htbp]
  \begin{center}
    \includegraphics[clip,width=10.0cm]{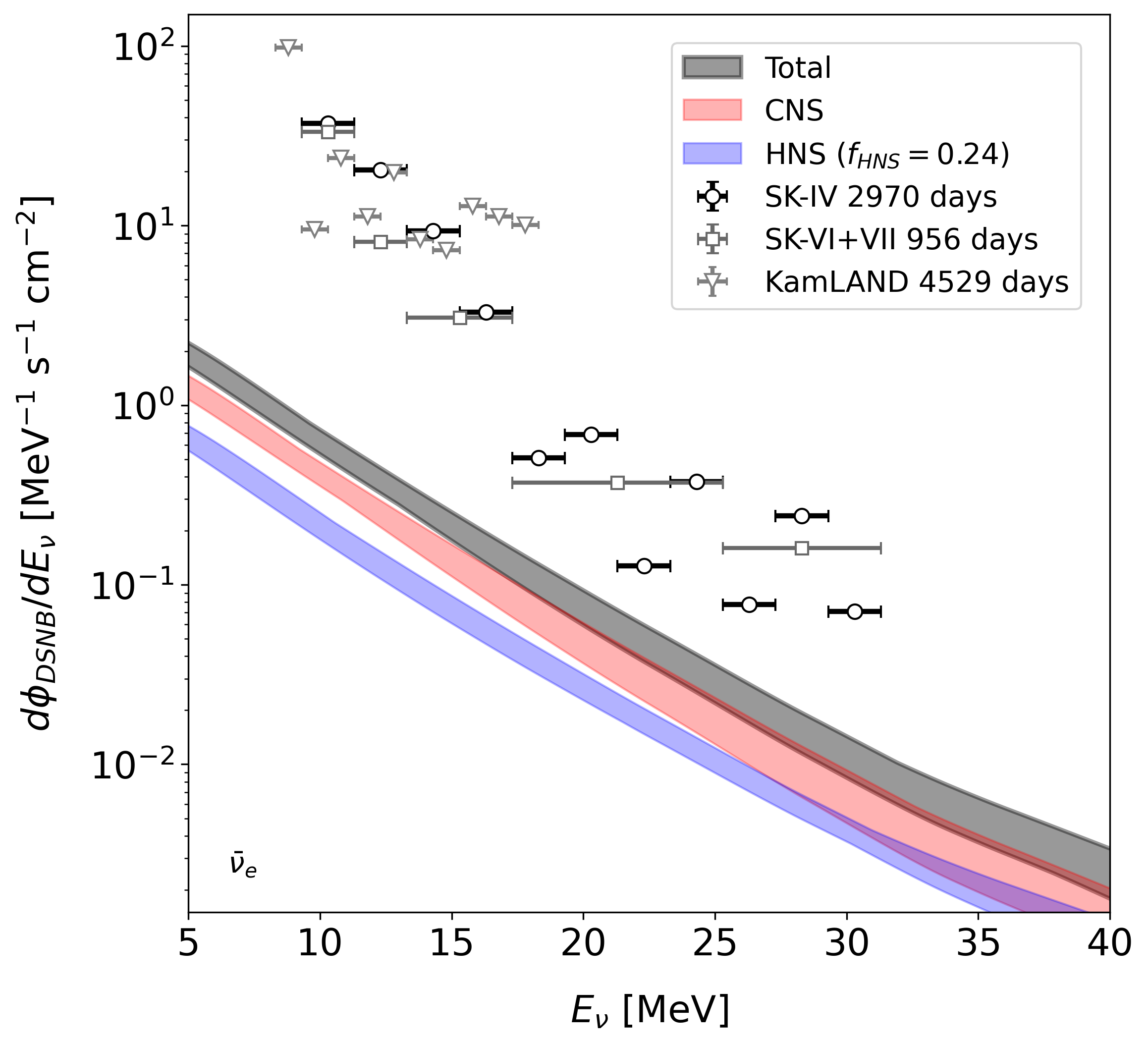}
  \end{center}
  \vspace{-10truept}
  \caption{Calculated DSNB $\bar{\nu}_e$ flux in comparison with experimental upper limits from Super-Kamiokande~\cite{2021PhRvD.104l2002A,2025arXiv251102222A} and KamLAND~\cite{2022ApJ...925...14A}. Contributions from the CNS and HNS cases with $f_{\rm HNS}=0.24$, and their sum are shown. The bands cover different choices of the nuclear EOS (LS220, Shen, or Togashi) and neutrino mass ordering (NMO or IMO).}
  \label{fig:dsnbflux}
  \end{figure}

In the energy regime of $O(10)$~MeV, the main detection channel of DSNB neutrinos is inverse beta decay of electron antineutrinos ($\bar{\nu}_e + p \rightarrow e^+ + n$)~\cite{2003PhLB..564...42S,2012RvMP...84.1307F,2022JHEP...08..212R}.
Hence, the DSNB flux of $\bar{\nu}_e$ is focused in this study. 
The $\bar{\nu}_e$ flux at Earth is a mixture of the fluxes of all flavors at production due to the matter effect~\cite{1978PhRvD..17.2369W,1985YaFiz..42.1441M}. 
The effect of this flavor mixing differs for neutrino mass ordering, {\it normal} (NMO) or {\it inverted} (IMO). 
Both NMO and IMO cases are considered in this study. 
The resulting $\bar{\nu}_e$ flux at Earth in each neutrino mass ordering case is as follows: 
  \begin{eqnarray}
   \frac{d\phi_{\bar{\nu}_e}}{dE_\nu} &\approx& 0.68 \times \frac{d\phi^0_{\bar{\nu}_e}}{dE_\nu} + 0.32 \times \frac{d\phi^0_{\bar{\nu}_x}}{dE_\nu} \ ({\rm NMO}), \\ 
   \frac{d\phi_{\bar{\nu}_e}}{dE_\nu} &\approx& \frac{d\phi^0_{\bar{\nu}_x}}{dE_\nu} \ ({\rm IMO}), 
  \label{eq:fluxobs}
  \end{eqnarray}
where $d\phi^0_{\bar{\nu}_e}/dE_\nu$ and $d\phi^0_{\bar{\nu}_x}/dE_\nu$ are the $\bar{\nu}_e$ and $\bar{\nu}_x$ ($= \bar{\nu}_\mu + \bar{\nu}_\tau$) fluxes at production, respectively. 
The calculated DSNB $\bar{\nu}_e$ fluxes at Earth for the CNS and HNS cases are shown in Figure~\ref{fig:dsnbflux}. 
Here, the fractional parameter of $f_{\rm HNS}=0.24$, a consequence of the mapping of SN types defined in Section~\ref{sec:snclass}, is used.
The results with different nuclear EOS and neutrino mass ordering cases are shown as bands. 
In the figure, the latest upper limits from Super-Kamiokande~\cite{2021PhRvD.104l2002A,2025arXiv251102222A} and KamLAND~\cite{2022ApJ...925...14A} are also shown. 
The current world record is away from the prediction in this study by a factor to $\sim$1 order of magnitude around $E_\nu \approx 20$~MeV. 

\subsection{Diffuse flux of non-thermal CSM-origin SN neutrinos} 

\begin{figure}[htbp]
  \begin{center}
    \includegraphics[clip,width=10.0cm]{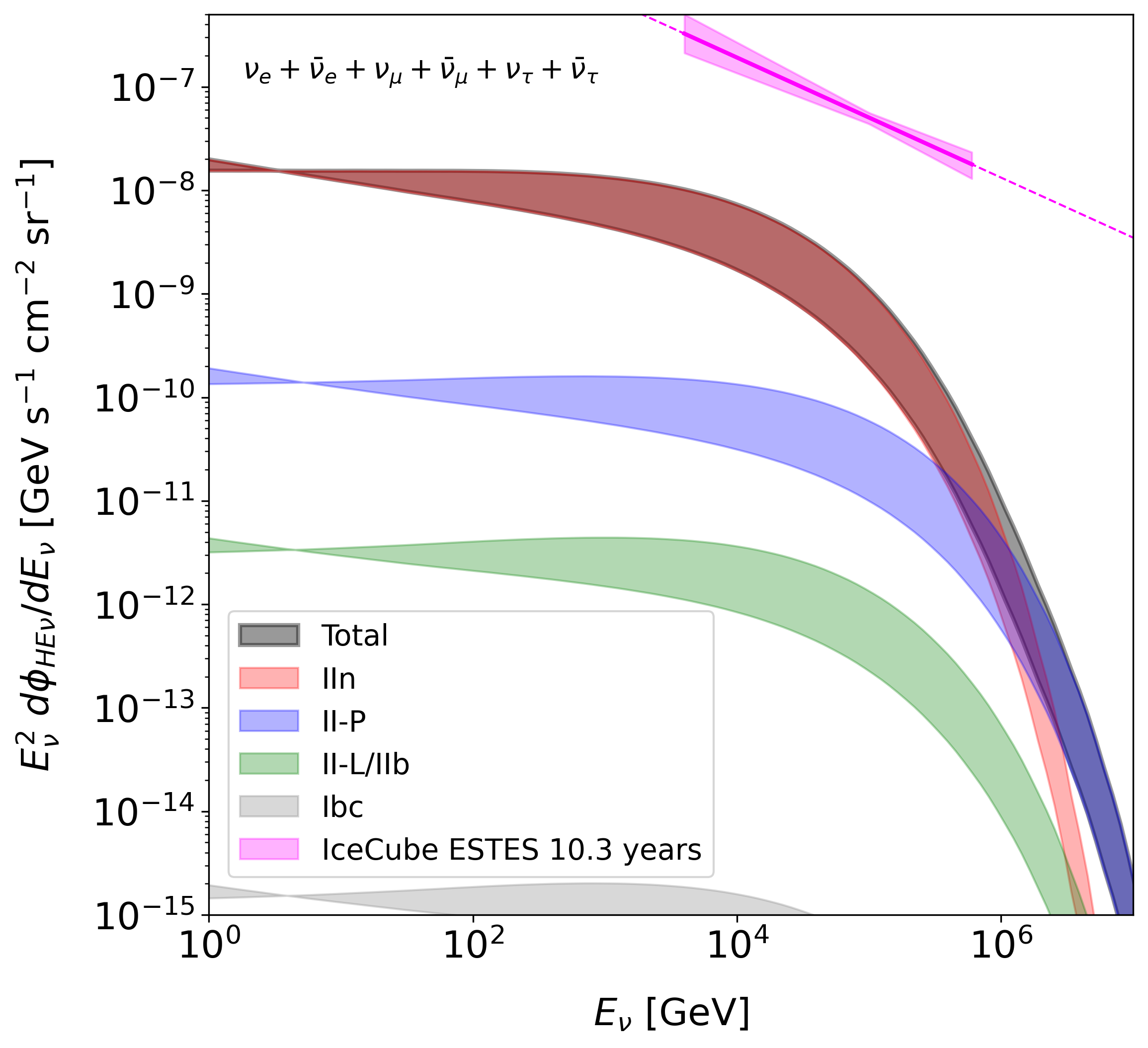}
  \end{center}
  \vspace{-10truept}
  \caption{Calculated diffuse high-energy SN neutrino flux ($\nu_e + \bar{\nu}_e + \nu_\mu + \bar{\nu}_\mu + \nu_\tau + \bar{\nu}_\tau$) in comparison with the measured diffuse astrophysical neutrino flux in a recent IceCube analysis using starting track events (ESTES)~\cite{2024PhRvD.110b2001A}. Contributions from each SN type and their sum are shown. The bands cover the different spectral indices of the parent cosmic ray flux ($2.0 \leq s \leq 2.2$).}
  \label{fig:diffhighenu}
  \end{figure}

The diffuse flux of SN high-energy neutrinos is calculated similarly to DSNB as:  
  \begin{eqnarray}
    \frac{d\phi_{{\rm HE}\nu}(E_\nu)}{dE_\nu}= \frac{c}{4\pi}\int^{z_{\rm max}}_{0} R_{\rm CC}(z) \left\langle \frac{dN_{\rm HE\nu}(E^\prime_\nu)}{dE^\prime_\nu} \right\rangle \nonumber \\ \times \frac{dz}{H_0\sqrt{\Omega_{\rm m}(1+z)^3+\Omega_\Lambda}},
  \label{eq:diffhenuflux}
  \end{eqnarray}
where $R_{\rm CC}(z)$ is the core-collapse rate calculated in Eq.~(\ref{eq:ccrate}). 
The same values are used for the cosmological parameters $\Omega_{\rm m}$, $\Omega_\Lambda$, $H_0$, and $z_{\rm max}$ as in Section~\ref{subsec:dsnb}. 
Different from the DSNB energy range, because attenuation of the high-energy neutrino flux due to Earth is significant in the context of experimental detection, the flux per solid angle is more useful. 
Therefore, a spherical integral over solid angles is not applied ($4\pi$ in the denominator is its reflection).  
The average neutrino number spectrum is calculated by summing that from each SN type ($dN_{\rm IIn}(E^\prime_\nu)/dE^\prime_\nu$, $dN_{\rm II\mathchar`-P}(E^\prime_\nu)/dE^\prime_\nu$, $dN_{\rm II\mathchar`-L/IIb}(E^\prime_\nu)/dE^\prime_\nu$, $dN_{\rm Ibc}(E^\prime_\nu)/dE^\prime_\nu$) with fractional parameters for each as: 
  \begin{eqnarray}
    \left\langle \frac{dN_{\rm HE\nu}(E^\prime_\nu)}{dE^\prime_\nu} \right\rangle &=& f_{\rm IIn}\frac{dN_{\rm IIn}(E^\prime_\nu)}{dE^\prime_\nu} + f_{\rm II\mathchar`-P}\frac{dN_{\rm II\mathchar`-P}(E^\prime_\nu)}{dE^\prime_\nu} \nonumber \\ &+& f_{\rm II\mathchar`-L/IIb}\frac{dN_{\rm II\mathchar`-L/IIb}(E^\prime_\nu)}{dE^\prime_\nu} + f_{\rm Ibc}\frac{dN_{\rm Ibc}(E^\prime_\nu)}{dE^\prime_\nu}, 
  \label{eq:nhenu}
  \end{eqnarray}
where $f_{\rm IIn}+f_{\rm II\mathchar`-P}+f_{\rm II\mathchar`-L/IIb}+f_{\rm Ibc}=1$. 
Here, the sum of all neutrino flavors ($\nu_e + \bar{\nu}_e + \nu_\mu + \bar{\nu}_\mu + \nu_\tau + \bar{\nu}_\tau$) is considered. 
The resulting diffuse flux is shown in Figure~\ref{fig:diffhighenu} together with the observed astrophysical neutrino flux from a recent analysis using 10.3 years of IceCube data~\cite{2024PhRvD.110b2001A} (the shown fluxes are multiplied by $E_{\nu}^2$ following the convention in the field).
Here, the results with $2.0 \leq s \leq 2.2$ as the spectral index of the original cosmic ray flux are shown as bands. 
The model prediction from this study can explain 5--10\% of the currently observed flux around $E_\nu \approx 10^4$~GeV. 

\section{Discussion} \label{sec:discuss}

As shown in Figures~\ref{fig:dsnbflux} and \ref{fig:diffhighenu}, there are some systematic factors that change the fluxes of thermal MeV and non-thermal high-energy neutrinos independently, e.g., the nuclear EOS, neutrino mass ordering, the CSM density profile, and the spectral index of parent cosmic rays in the current study. 
Therefore, detecting the diffuse neutrino flux at multiple energies enables an access to different aspects of CCSNe, which will give rise to a comprehensive understanding of collapsing massive stars. 
It should also be mentioned that the current predictions are strongly tied with the core-collapse rate inferred from optical survey; however, neutrinos might have been produced from collapsed stars that were not captured by such survey for some reason. 
The BH formation, as described below, is one of such cases. 

Not only independent systematic factors, but also there are many more common factors that change the DSNB and high-energy SN neutrino fluxes. 
SFRD is among the most influential factors as its choice is directly reflected as a scale of the diffuse flux. 
In Refs.~\cite{2023ApJ...950...29A,2024PhRvD.109b3024E}, the SFRD impact on the DSNB flux is discussed, where the systematic effect is estimated to be as large as 10--30\%. 
A similar size effect is expected for the diffuse flux of high-energy SN neutrinos. 
IMF is another important choice which determines the core-collapse rate as seen in Eq.~(\ref{eq:ccrate}).
Systematic impacts of the IMF form are discussed in many articles (e.g., see Ref.~\cite{2022MNRAS.517.2471Z}). 
Note that switching the IMF form may require mapping of the SN type along progenitor mass proposed in Ref.~\cite{2011MNRAS.412.1522S}. 
In addition, the progenitor fate influenced by binary interactions is recently investigated in a theoretical way~\cite{2021PhRvD.103d3003H,2024MNRAS.532.3926K}. 
Further progress in such area may advance categorization of the SN type. 
 
In Figures~\ref{fig:dsnbflux} and \ref{fig:diffhighenu}, the BH formation is ignored, but there should be some contributions from such case~\cite{2024arXiv241014778D,2025OJAp....8E.167S}. 
A recent search for failed SNe in nearby galaxies shows a fraction of failed SNe to be 4--39\% at a 90\% confidence level~\cite{2021arXiv210403318N}. 
On another hand, in Ref.~\cite{2020ApJ...896...56W}, theoretical prediction on the BH fraction is given as 9--32\%.
In addition, a failed SN candidate was recently claimed in the Andromeda Galaxy~\cite{2024arXiv241014778D}. 
A choice of the BH formation fraction ($f_{\rm BH}$) changes the CCSN rate as $(1-f_{\rm BH}) \times R_{\rm CC}(z)$. 
This may decrease the non-thermal high-energy SN neutrino flux, while the DSNB flux is enhanced at higher energies, especially at $E_\nu \gtrsim 30$~MeV, not only is decreased at lower energies contributed mainly from the CNS and HNS cases.
It should be noted that inclusion of failed SNe in the scheme would change mapping of the SN types on the Salpeter IMF in the high mass region hence impact the resulting diffuse fluxes. 
The systematic effects on DSNB from the failed SN fraction are investigated in other studies~\cite{2022ApJ...937...30A,2023ApJ...953..151A,2024PhRvD.109b3024E,2024ApJ...975...71N}.

As mentioned before, a complete modeling of the entire process from core collapse to ejecta-CSM interactions is challenging and not yet performed; however, a bunch of theoretical studies have made huge progress in each physical stage (e.g., see Refs.~\cite{2018SSRv..214...31T,2021Natur.589...29B} and references therein). 
Towards such comprehensive understanding, a long-term modeling of the core-collapse process is essential so that the outcomes of explosions may be available as reliable inputs for the later process that produces high-energy neutrinos. 
Recently, the late-time phase after core collapse, including proto-NS cooling and fallback mass accretion, are studied in many ways~\cite{2021PhRvD.103b3016L,2021PTEP.2021b3E01M,2022PhRvD.106d3026E,2024ApJ...960..116A}. 
At the same time, because CSM is thought to be formed ahead of the core-collapse process, detailed understanding of stellar evolution of single and binary stars before explosion is also essential~\cite{2011MNRAS.415..199M,2014A&A...564A..83M,2018NatAs...2..808F,2021MNRAS.503.2432B,2024MNRAS.532.3926K,2025ApJ...982...93S}.

Prospects for detecting the diffuse fluxes of SN neutrinos are promising as larger volume and higher performance neutrino telescopes are coming. 
For the low-energy neutrinos (DSNB), Hyper-Kamiokande~\cite{2018arXiv180504163H} is planned as a successor of the on-going Super-Kamiokande detector with a $\sim$10 times larger volume and JUNO has just started its operation in China with the best sensitivity expected at lowest energies ($E_\nu \lesssim 10$~MeV)~\cite{2022Univ....8..181L,2022JCAP...10..033A,2023arXiv231116550C,2025arXiv251114590A}. 
While the astrophysical neutrino flux is detected at IceCube, further precise measurements at IceCube-Gen2~\cite{2021JPhG...48f0501A} and KM3NeT~\cite{2016JPhG...43h4001A} are expected to help elucidate the neutrino sources, providing some insights into CCSN-origin high-energy neutrinos. 
These future experiments will play important roles in detecting transient SN sources as well~\cite{2020PhRvD.101l3018H,2021ApJ...916...15A,2021EPJC...81..445A,2023ApJ...945...98V,2023ApJ...956L...8K}. 
Combining transient and diffuse neutrino measurements will dramatically promote the understanding of SN properties and corresponding theoretical modeling.

\section{Conclusion} \label{sec:concl}

In this paper, diffuse SN neutrino fluxes in two different energy ranges are forcibly linked based on the assumptions motivated by optical survey, and the importance of detecting such diffuse neutrino flux at multiple detectors is discussed.
A conventional SN classification is adopted and the observationally motivated mapping of each type onto the Salpeter IMF for the mass range of 8.5--150$M_\odot$ is referred to bridge between two energy regimes. 
Models of thermal SN neutrinos based on a progenitor with $15 M_\odot$ and $40 M_\odot$ are assigned to SNe lighter and heavier than $23.1 M_\odot$ on the Salpeter IMF, respectively. 
In this scheme, SNe Type II-P, II-L, and partial IIb/Ibc are assumed to arise from lighter progenitors, and Type IIn and partial IIb/Ibc are from heavier ones.
Models of neutrino production from ejecta-CSM interactions are taken for high-energy SN neutrinos with different profiles of SN ejecta and CSM being assumed for each SN type (IIn, II-P, II-L/IIb, and Ibc). 
Both thermal MeV-scale and non-thermal high-energy neutrino fluxes are convoluted with the common cosmological constants and CCSN rate based on the Illustris-1 cosmological simulation and the Chabrier IMF to serve the diffuse fluxes at Earth. 
The resulting DSNB $\bar{\nu}_e$ flux is close to the current upper limit from Super-Kamiokande within a factor to 1 order of magnitude. 
The calculated diffuse flux of high-energy SN neutrinos can explain up to $\sim$10\% of the currently observed flux of astrophysical neutrinos at IceCube. 
Note that these fluxes could be further increased or decreased by other systematic factors as discussed in Section~\ref{sec:discuss}. 
This work, even though still at a conceptual and immature phase, provides a new multifaceted perspective on studying CCSNe via neutrino detection at different energies. 
Further progress in theoretical studies on the stellar evolution as well as core-collapse process will realize a scheme that unifies two physical processes in a more consistent way.

\acknowledgments

The author thanks Ken'ichiro Nakazato, Kohta Murase, and Ryo Sawada for fruitful discussions.
This work was partially supported by NSF Grant No. PHY-2309967.

\bibliographystyle{apsrev4-2}
\bibliography{reference}  

\end{document}